\documentclass[a4paper]{jpconf}
\usepackage{graphicx}%
\usepackage{latexsym}
\usepackage{amssymb}

\newcommand{\vj}{{\mathbf{j}}}
\newcommand{\vA}{{\mathbf{A}}}
\newcommand{\vE}{{\mathbf{E}}}
\newcommand{\vG}{{\mathbf{G}}}
\newcommand{\vp}{{\mathbf{p}}}
\newcommand{\vrr}{{\mathbf{r}}}

\begin{document}
\title{Photoemission Beyond the Sudden Approximation}
\author{Carl-Olof Almbladh}
\address{Solid State Theory, Department of Physics, Lund
  University, Sweden}
\date{\today}

\begin{abstract}
The many-body theory of photoemission in solids is reviewed with
emphasis on methods based on response theory. The classification
of diagrams into loss and no-loss diagrams is discussed and related
to Keldysh path-ordering book-keeping. Some new results on
energy losses in valence-electron photoemission from free-electron-like
metal surfaces are presented. A way to group diagrams is presented
in which spectral intensities acquire a Golden-Rule-like form
which guarantees positiveness. This way of regrouping should be
useful also in other problems involving spectral intensities, such as
the problem of improving the one-electron spectral function away
from the quasiparticle peak.

\end{abstract}

\section{Introduction}
\label{sec:intro}
Photoemission spectroscopy\cite{einstein} (PES) is since long
a major tool for gaining information of the electronic
structure of matter.  
For instance, angular resolved photoemission spectroscopy (ARPES)
makes it possible to measure the occupied quasiparticle bands,
and photoemission from core levels yields important local
information at the atom being excited. Owing to the usually
rather short mean free path, PES is also well suited for studying surface 
states, 
adsorbates at surfaces, etc.

Photoelectron spectra are usually interpreted
in terms of the one-electron spectral function $A(\epsilon)$ corresponding
to a sudden removal of an electron. In emission from solids,
however, the photoelectron is created at a certain distance from
the surface and may undergo losses on its way out. 
In the sixties, Berglund and Spicer\cite{spicer} proposed a 
semi-empirical model involving primary
excitation, transport to the surface, and transmission through the
surfaces as three separate steps. Clearly a three-step description
can at best be an approximation. An important step forward was taken
by Schaich and Ashcroft\cite{schaich} and by Mahan\cite{mahan}, 
who formulated the problem as a one-step quantum-mechanical 
process and presented model results for independent electrons.
The Schaich-Ashcroft response formulation was soon generalized to
account for interaction by Caroli \textit{et al.}\cite{caroli} 
who applied the formalism to phonon effects.
Among important early works I would also like to mention the
studies of plasmon losses by Langreth and collaborators
\cite{lang, chang-lang1, chang-lang2, lang2}. In particular,
they explain why the satellites are greatly modified and
suppressed also when the excitation energy is quite high,
and why the satellites are so weak in, say, x-ray emission.

In strict one-electron theory, the photoelectrons cannot undergo
losses by scattering against other electrons, phonons, and impurities.
As a consequence, the entire solid will contribute to the photocurrent
which is then limited by the penetration depth of the electromagnetic
radiation. In reality, the current is limited by the escape depth
of the photoelectrons, in practice usually just a few atomic layers.
Thus the mean free path must somehow enter the theory, and 
in the mid-seventies ways to remedy this shortcoming were developed
by Feibelman and Eastman\cite{feibelman}, Pendry\cite{pendry}, 
and Liebsch\cite{liebsch,liebsch2}.
In essence, the no-loss photocurrent is to be calculated in a one-step
but independent-electron way with one important exception: The 
photoelectron is assumed to move in the non-Hermitian
optical potential, the (advanced) self-energy $\Sigma_a$. 
The photoelectron
orbitals then get damped inside the sample, and the yield becomes
limited by the photoelectron escape depth as it should. Today there
exists several computer codes based on these ideas which take
the underlying bandstructure into account in a realistic way.

The crucial
point of introducing damped photoelectron orbitals was first
obtained by Langreth\cite{lang2} in the case of impurity scattering,
and by Caroli \textit{et al.}\cite{caroli} in the case
when the escape depth is limited by phonons. In the mid-eighties,
by Almbladh\cite{alm85} and Bardyszewski and Hedin
\cite{hedin85} considered the problem of interacting electrons using two 
seemingly rather different techniques, one based
on many-body perturbation theory and one based on scattering theory.
These works largely justify the above one-electron-like methods for
the no-loss photocurrent but they also predict corrections, which, for instance,
modify the optical matrix elements.

There are several excellent recent
reviews of solid-state photoemission \cite{schattkebook,hufner,kevan}. 
In this paper, I would
like to concentrate on how the techniques of non-equilibrium 
Green's functions can be used both for obtaining approximations
suitable for calculations in real materials as well as for obtaining
general results like the structure of the quasiparticle no-loss part
of a photoelectron spectrum.

\section{Basic theory}
\label{basic}
I begin by a brief account of the quantum-mechanical
description of the steady photoelectron current. As mentioned above,
methods based on scattering theory\cite{adawi,mahan,hedin85}, and methods 
based on quadratic response formalism and many-body perturbation theory\cite{schaich,caroli,
hermeking,lang,lang2,alm85}. I begin by one-particle theory in
order to illustrate the various limiting procedures, and generalize
then to interacting systems.

\subsection{Independent particles}
\label{sec:oneparticle}
Let us first consider independent electrons.
The photoemission process can be considered as an inelastic scattering process
with a photon in the initial state and an asymptotically free electron
in the final state. In order to strengthen the analogy with scattering,
let us take the system to occupy an finite volume $v$ enclosed
in a quantization box of volume $V$. Because the photoelectron enters
in the final rather than the initial state, its wavefunctions will obey ingoing
wave boundary conditions\cite{gell-mann,adawi}. To first order in the radiative
coupling we obtain
\begin{equation}
\label{eq:i0scatt}
J_p = 2 \pi \sum_n^{occ} | \langle \chi_p^- | \Delta | n \rangle |^2 
\delta(\epsilon_p - \omega - \epsilon_n) .
\end{equation}
Here, $J_p$ is the momentum-decomposed photocurrent,
$\Delta$ the coupling to the radiation field (\textit{cf}. Eq. (\ref{eq:dH})), 
$\omega$ the photon energy,
and $\epsilon_\nu$ one-electron energies. Atomic (Hartree) units are used.

The photoelectron spectrum probes the occupied part of the one-electron
spectrum. For independent electrons, the spectral function can be written
\begin{equation}
A(\epsilon) = \sum_n | \phi_n \rangle \delta(\epsilon - \epsilon_n) 
\langle \phi_n |
\end{equation}
in terms of eigen-orbitals $\phi_n$. In terms of the spectral function,
we have
\begin{equation}
J_p = 2 \pi \langle \chi_p^-  | \Delta A(\epsilon_p - \omega) \Delta^\dag |
\chi_p^-  \rangle 
\end{equation}
for photoelectron energies below threshold $\mu + \omega$. Here $\mu$ is
the Fermi level relative to the vacuum level.

In order to compare with the response formulation of Schaich and Ashcroft,
let us introduce orbitals $\phi_{\tilde{p}}$ which are plane waves
well outside the sample and which vanish inside, and let $c_{\tilde{p}}$
be the associated destruction operator.
We assume that the external field has been slowly switched on
in the remote past, (modulated by $e^{\eta t }$), and take the average
$2 \eta \langle c_{\tilde{p}}^+ c_{\tilde{p}} \rangle$ as a measure
of the steady momentum-decomposed photocurrent. The details how the plane
waves are truncated inside and near the sample will not influence
the final results.
We have
\begin{equation}
\label{eq:kt}
K(t) = H + \left [ \delta H e^{-i\omega t }   + \delta H^\dag e^{i\omega t }
\right ] e^{\eta t} ,
\end{equation}
where $H$ describes the system in the absence of external fields, and where
\begin{equation}
\label{eq:dH}
\delta H = - \int \vj(\vrr) \cdot \vA(\vrr) = 
\sum_{kl} c_k^+ c_l \Delta_{kl} 
\end{equation}
gives the coupling to the radiation field.
We use a radiation gauge such that the scalar potential $\delta \Phi$ vanishes. 
(The diamagnetic
coupling then gives no contribution to leading order, see Ref \cite{schaich}.)
We assume that the system has been subject to the field for $t<0$,
and define the photocurrent by
\begin{equation}
J_p   = \lim_{\eta \rightarrow 0^+} 
2 \eta \langle c^+_{\tilde{p}} c_{\tilde{p}} \rangle .
\end{equation}
As in scattering theory the infinite volume limit should be taken before 
the limit $\eta \rightarrow 0$ , 
i.e., $\eta$ should be considered large compared to the 
level spacing due to the quantisation box V.
Following Liebsch\cite{liebsch}, we develop each initial-state
orbital $n$ in time to obtain
\begin{equation}
\phi_{n,\eta}(t=0) =   \left [
1  + G^r(\epsilon_n + \omega  + i \eta) \Delta 
+ G^r(\epsilon_n - \omega  + i \eta) \Delta^\dag \right ] \phi_n ,
\end{equation}
where $G^{r/a}(\epsilon)  = 1 / (\epsilon - h \pm i \delta) $ are the
retarded/advanced one-electron Green's functions, and $h_{kl}$
the one-electron Hamiltonian matrix of the unperturbed
system.  We next form the expectation value
$\langle c^+_{\tilde{p}} c_{\tilde{p}} \rangle $
and sum over
occupied states initial states to obtain
\begin{eqnarray}
\label{eq:jp1}
J_p  &=& \lim 2 \eta \sum_n^{occ}
| \langle \tilde{p} | 
G^r(\epsilon_n + \omega  + i \eta) \Delta | n \rangle |^2 \\
&=& \lim_{\eta \leftarrow 0}
2 \eta \int_{- \infty}^0 dt \int_{- \infty}^0 dt^\prime
\; i \; \langle \tilde{p} | G^r(-t^\prime) \Delta G^<(t^\prime - t)
\Delta^\dag G^a(t) | \tilde{p} \rangle \;
e^{i \omega (t - t^\prime)} \; e^{\eta (t + t^\prime)} .
\label{eq:jp2}
\end{eqnarray}
Here, 
$G^<$ is the usual lesser function, 
$\langle k | G^< (t)| l \rangle = i \langle N | c_k^+(0) c_l(t) | N \rangle$,
and $| N \rangle$ the $N$-electron ground state.
In order to obtain Eqs. (\ref{eq:jp1}, \ref{eq:jp2}) we have used the fact 
that electrons in initial states $n$ are confined to the solid, 
and we have dropped the negative-frequency term
which is easily seen not to contribute to the photocurrent in
the limit $\eta \rightarrow 0^+$. 
In this limit one can further replace the truncated orbital
$\phi_{\tilde{p}}$ with a plane wave $\phi_p$  (see \textit{e.g} Ref. 
\cite{alm85}).  Eq. (\ref{eq:jp2}) is then equivalent to the Schaich-Ashcroft 
independent-electron result.

In order to see the equivalence between the scattering and response
formulations, we express $G^{r/a}$ in terms of the Green's function
$G^{r/a}_0$ for free space and the scattering matrix $T^{r/a}$,
\begin{equation}
G^{r/a}(\epsilon) = G^{r/a}_0(\epsilon) + G^{r/a}_0(\epsilon) 
T^{r/a}(\epsilon) G^{r/a}_0(\epsilon) .
\end{equation}
As remarked above, we can replace the truncated plane wave $\phi_{\tilde{p}}$
with $\phi_p$ in the limit $V \rightarrow \infty$, $\eta \rightarrow 0^+$.
In this limit we have
\begin{eqnarray}
\langle n | 1 &+& \Delta G^a(\epsilon_n + \omega - i \eta) | \tilde{p} \rangle
=
\langle n | \Delta G^a(\epsilon_n + \omega - i \eta) | \tilde{p} \rangle
\nonumber \\
&=& \frac{1}{\epsilon_n + \omega - \epsilon_p - i \eta}
\langle n | \Delta [ 1 + G^a_0(\epsilon_p - i \eta) T^a(\epsilon_p - i \eta) ]
| p \rangle \nonumber
\end{eqnarray}
and the wave operator $1 + G^a_0 T^a$ transforms $\phi_p$
into  $\chi^-_p$. In this way, 
\begin{displaymath}
| \langle \tilde{p} | G^r(\epsilon_n + \omega  + i \eta) \Delta | 
n \rangle |^2 
\rightarrow \frac{2\eta \langle \chi^-_p | \Delta | n \rangle|^2}
{(\epsilon_n + \omega - \epsilon_p)^2 + \eta^2}
\rightarrow 2 \pi | \langle \chi^-_p | \Delta | n \rangle|^2
\delta((\epsilon_n + \omega - \epsilon_p) ,
\end{displaymath}
which establishes the equivalence between the response and the
scattering approaches for independent electrons.

The geometry with a finite but possibly macroscopic
sample is convenient for performing the somewhat
subtle limits above, but in actual calculations one
usually let the sample grow until it fills a half-space.
The final-state orbitals $\chi^-_p$ are then replaced by 
their two-dimensional (2D) analogues, \textit{i.e.}, time-reversed
states for scattering against the surface of the
sample. These states are usually referred to as time-reversed
LEED states because similar 2D scattering enters in the problem of low
energy electron diffraction (LEED).

As stressed in the Introduction, strict one-particle theory
is unphysical and leads to a yield limited by the radiation
penetration depth rather than the photoelectron escape depth.
If the propagators in Eq. (\ref{eq:jp2}) are replaced by their
interacting counterparts, the final-state orbitals get damped 
inside the sample and the photocurrent limited by the escape 
depth as it should.

\subsection{Interacting electrons}
\label{interacting}
\begin{figure}[htbp]
\begin{center}
\includegraphics*[scale=0.6]{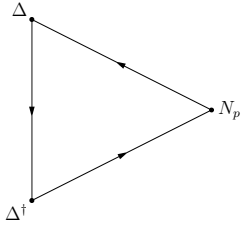}
\caption{The lowest-order non-vanishing diagram for the 
photoemission process}
\label{pe0}
\end{center}
\end{figure}

We expand $\langle c_{\tilde{p}}^+  c_{\tilde{p}} \rangle$
directly in orders of the perturbation.
The linear response vanishes identically because all electrons
are confined to the sample in the ground state, 
$\tilde{c}^+_p \tilde{c}_p | N \rangle = 0$. In the radiation
gauge, the diamagnetic coupling involving $\vA^2$ gives no
contribution. The remaining contributions can be written
\begin{equation}
\label{eq:jpi1}
J_p(\epsilon)  = \lim_{\eta \rightarrow 0^+}
2 \eta \int_{- \infty}^0 dt \int_{- \infty}^0 dt^\prime
e^{i \omega (t - t^\prime)} \; e^{\eta (t + t^\prime)} 
\langle N | \delta H^\dag(t) \hat{N}_p(0) \delta H(t^\prime) 
| N \rangle .
\end{equation}
Eq. (\ref{eq:jpi1}) involves a three-particle path-ordered
Green's function
\begin{equation}
\label{eq:Rp}
R_p(t, t^\prime) =
\sum_{klmn} 
\Delta^*_{k,n}\Delta_{l,m}
\langle N | c^+_k(t) c_n(t) c^+_{\tilde{p}}(0)c_{\tilde{p}}(0)
c^+_l(t^\prime) c_m(t^\prime) | N \rangle ,
\end{equation}
which may be evaluated by standard Keldysh technique\cite{keldysh}.
To lowest order, all unconnected contractions vanish because
\begin{equation}
G^<(t) | \tilde{p} \rangle = 0, \; \langle \tilde{p} | G^< (t) = 0
\label{eq:constraint}
\end{equation}
both with and without interaction.
By the same argument, there is only one non-vanishing connected
contraction (see Fig. \ref{pe0}), which gives
\begin{eqnarray}
R^{(0)}_p(t, t^\prime) &=& 
i \langle \tilde{p} | G^>_0(-t^\prime) \Delta G^<_0(t-t^\prime) 
\Delta^\dag G^>_0(t) | \tilde{p} \rangle \nonumber \\
&=&
i \langle \tilde{p} | G^r_0(-t^\prime) \Delta G^<_0(t-t^\prime) 
\Delta^\dag G^a_0(t) | \tilde{p} \rangle 
\label{eq:jp0k}
\end{eqnarray}
when $t, \; t^\prime < 0$.
Eq. (\ref{eq:jp0k}) gives back
the previous independent-electron result in Eq. (\ref{eq:jp2}).

\begin{figure}[htbp]
\begin{center}
\includegraphics*[scale=0.6]{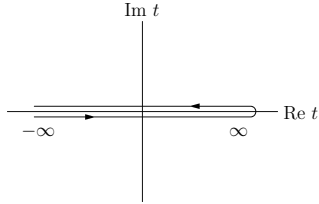}
\caption{Keldysh contour at $T=0$}
\label{contour}
\end{center}
\end{figure}

In the present problem, the correlation 
$R_p(t, t^\prime)$ in Eq. (\ref{eq:Rp}) is a ground-state
correlation which simplifies the book-keeping of Keldysh diagrams.
The Keldysh propagators refer to the contour in Fig. (\ref{contour}) 
which means
that we assume that the interacting ground state can be generated
by switching on the interaction adiabatically in the remote past,
\begin{equation}
R_p(t, t^\prime) =
\langle N, 0 | S^\dag(\infty) \delta H^\dag(t) c^+_{\tilde{p}} S(\infty)
 S^\dag(\infty) c_{\tilde{p}} \delta H(t^\prime) S(\infty)
| N,0 \rangle .
\end{equation}
Thus $t = t_-$, $t^\prime = t^\prime_+$, and $c^+_{\tilde{p}} c_{\tilde{p}}$
is straddling the two branches of the Keldysh contour in Fig. (\ref{contour}).
(The indices $\pm$ refer to the positive and negative time-ordered parts, 
respectively.)
The propagators joining points $\mu,\; \nu$ of the different
time-ordered parts are
\begin{displaymath}
G(t_\mu, t_\nu) = 
\begin{array}{l}
G^c(t - t^\prime),\;\; \mu,\;\nu = + \\
G^{\tilde{c}}(t - t^\prime),\;\; \mu,\;\nu = - \\
G^<(t - t^\prime),\;\; \mu = +, \;\nu = - \\
G^>(t - t^\prime),\;\; \mu = -, \;\nu = + \;,
\end{array}
\end{displaymath}
where $G^{c}$, $G^{\tilde{c}}$ are propagators with positive
and negative time-ordering, respectively, and $G^{\lessgtr}$ the usual
lesser/greater functions. At zero temperature we have
\begin{equation}
a^\lessgtr(\epsilon) b^\gtrless(\epsilon) = 0
\end{equation}
for any Fermi propagators $a$ and $b$. Further, with the above assumption
of adiabatic switching, the expansion of a time-ordered propagator
will only involve propagators with the same time ordering.

\begin{figure}[htbp]
\begin{center}
\includegraphics*[scale=0.6]{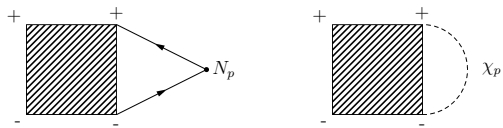}
\caption{Reduction of a general photoemission diagram}
\label{p_gen}
\end{center}
\end{figure}

The contribution from a  general diagram
can be reduced to a form analogous to the one-electron Golden Rule
form in Eq. (\ref{eq:i0scatt}).
In the expansion in full Green's functions, any diagram
ends with a pair of electron lines joined to the
vertex $N_p$ representing the analyser (Fig. \ref{p_gen}). This means that
the correlation $R_p$ in Eq. (\ref{eq:Rp}) is of
the general form
\begin{equation}
R_p(t, t^\prime) = - \int \; dx \; d\tau \int \; dx' \; d\tau'
\langle \tilde{p} | G^c(-\tau') | x' \rangle
\langle x' | I(t, t'; \tau , \tau') | x \rangle
\langle x | G^{\tilde{c}}(\tau) | \tilde{p} \rangle ,
\end{equation}
where $I$,
the remain when the exit fermion lines have been removed,
is invariant under a simultaneous translation of
all four times and can be written
\begin{equation}
I(t, t'; \tau , \tau') =
\int \frac{d\nu d\nu' d\nu''}{(2 \pi)^3}
I(\nu, \nu'; \nu'',\nu' + \nu'' - \nu) \;
e^{-i\nu t + i \nu' t'} e^{i \nu'' \tau - (\nu + \nu' - \nu'')\tau'} .
\end{equation}
We notice
that $G^c$ and $-G^{\tilde{c}}$ can be replaced by retarded and advanced
propagators, respectively, since $\tilde{p}$ projects out the particle
part. The action of $G^{r/a}$ on the orbitals $\tilde{p}$ and the limit
$\eta \rightarrow 0$ can now be done analytically in much the same
way as for independent particle, which gives
\begin{equation}
\label{eq:JpI}
J_p = \langle \chi^-_p | I(\omega, \omega; \epsilon_p, \epsilon_p)
| \chi^-_p \rangle.
\end{equation}
Here $\chi^-_p$ is an one-electron time-reversed LEED orbital solved in 
the potential $V_C + \Sigma^a(\epsilon_p)$. ($V_C$ is the total 
Coulomb potential from the nuclei and the equilibrium electron density.) 
The self-energy $\Sigma^a(\epsilon)$
is non-Hermitian with a finite imaginary part inside the sample.
This makes $\chi^-_p$ decaying inside the sample with a decay length determined
by energy losses. Thus all parts of the spectrum including the no-loss
part is now limited by the escape depth of the photoelectrons.

In the case of optical responses and transport, the ideas of
self-consistent approximations put forward by Kadanoff and 
Baym\cite{kadbaym1, kadbaym2, baym} has been extremely
successful. It is the present author's opinion that this level
in not possible to achieve in the case of photoemission. In particular,
the Bethe-Salpeter equation which emerges as a result of selfconsistency
will include interaction lines of all orders. If we open up such
a diagram corresponding to a given approximation 
to the self-energy, infinitely many photoemission
diagrams will result. For sp-bonded solids, the
most important conserving approximation is
the self-consistent screened exchange approximation\cite{baym,hedin65}, usually
referred to as the $GW$ approximation.
Fortunately, the fulfilment of macroscopic
conservation laws are less crucial for useful approximations 
to photoelectron spectra. In photoemission, only a small fraction of
the emitted photoelectrons are actually measured. The majority
of photoelectrons which make up the overall background is usually
of no interest at all; instead the interest is focused on the spectral shapes
in a limited energy window. Often, processes of different
orders will contribute at different energies. Losses to plasmons
is here a typical example. What is important is to have approximations
which can explain the observed spectral profiles and 
which fulfil basic requirements
such as positiveness.

\subsection{Low-order diagrams}
\begin{figure}
\begin{center}
\includegraphics*[scale=0.7]{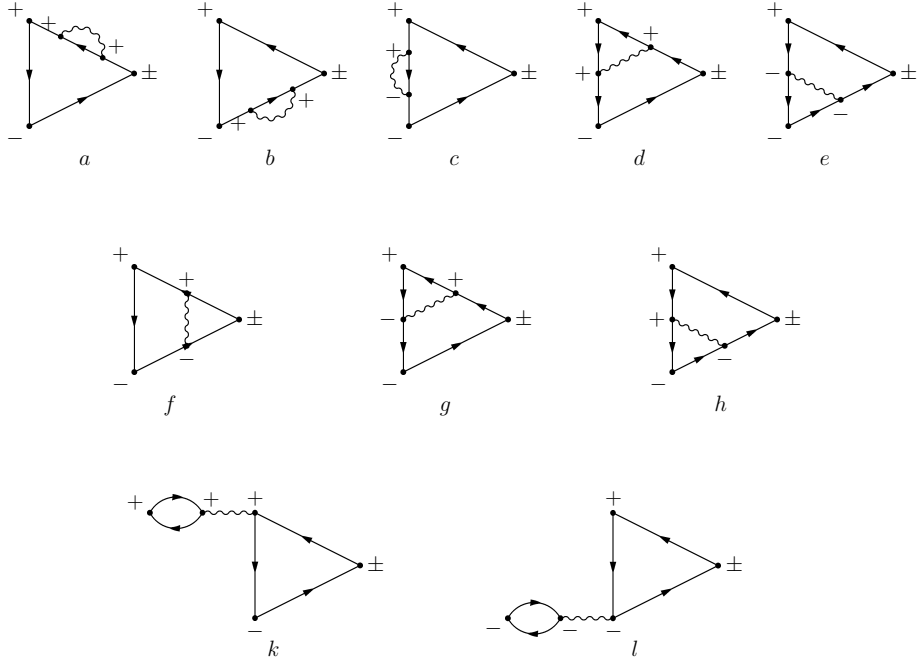}
\caption{First-order diagrams for the photoemission process}
\label{pe1}
\end{center}
\end{figure}

Let us expand the basic 3-particle correlation $R_p$ 
in orders of the screened interaction 
$W(\omega)=v \epsilon^{-1}(\omega)$.
The non-vanishing first-order diagrams are shown in Fig. (\ref{pe1}).
As was the case for independent electrons, the condition in 
Eq. (\ref{eq:constraint}) limits the number of non-vanishing contractions.
Vertices belonging to the forward (backward) time-ordered parts are
are labelled + (-), respectively.
The diagrams (a-c) involve
only self-energy insertions and should be omitted if we expand
in the full Green's function.
The remaining first-order diagrams fall into two classes,
namely those with interaction line joining point in the same
leg (d and e) and those with the interaction line joining
points in different legs (f, g, and h).
Consider first diagram f. It is well known that this diagram
describes an extrinsic loss, but to illustrate the
technique we give some details of the analysis.
The interaction line here represents the boson-like correlation
\begin{equation}
 W_> (\vrr t ,\vrr' t') = -i\langle N|\delta V_H (\vrr t) \delta V_H ( \vrr' t')
 | N \rangle
 = -i \int_0^\infty \; d \omega^\prime B(\vrr, \vrr^\prime, \omega^\prime )
 e^{-i \omega^\prime ( t - t^\prime)},
\end{equation}
where the spectral function $B$ is given by the dynamical structure
factor $S(\vrr,\vrr',\omega)$ convoluted with two Coulomb interactions
($v$), $B(\omega)=v*S(\omega)*v$. The fermion line from t' to t represents
$G^<$.
and the remaining lines time-ordered or anti-time-ordered functions 
$G^{c/\tilde{c}}$. After reduction according to Eq. (\ref{eq:JpI}) we obtain
\begin{equation}
\label{eq:loss_f}
J_p^{(f)} = 2 \pi \int_{-\infty}^0 d\epsilon \; \int_0^\infty \; d\omega^\prime
\delta(\epsilon + \omega - \omega^\prime - \epsilon_p)
\langle \chi_p^- | \underbrace{G^c(\epsilon + \omega) \Delta A(\epsilon) 
\Delta^\dag G^{c\dag}
(\epsilon + \omega)}_{B(\omega^\prime)} | \chi_p^- \rangle 
\end{equation}
where we have used an underbrace to indicate
how the spatial coordinates in $B(\omega)$ should be connected to the
electron propagators. Here and in the following we use the identity
$G^{\tilde{c}}(\epsilon) = - G^{c \dag}(\epsilon)$ in order to express
final results in spectral or lesser functions and functions with 
positive time-ordering.

The obvious physical interpretation of Eq. (\ref{eq:loss_f}) is that it describes
primary photo-excitation from an initial occupied level ($\epsilon$) to a
level of energy $\epsilon + \omega$. The photoelectron then suffers an
energy loss by exciting a plasmon or electron-hole pair of energy
$\omega^\prime$ and hits the detector with an energy 
$\epsilon + \omega - \omega^\prime$. Consider next
the diagrams g and h in Fig. 3. Using a similar technique as above
we obtain
\begin{equation}
\label{eq:loss_gh}
J_p^{(g)} + J_p^{(h)} = 4 \pi \Re \int_{-\infty}^0 d\epsilon \; \int_0^\infty \; d\omega^\prime
\delta(\epsilon + \omega - \omega^\prime - \epsilon_p)
\langle \chi_p^- | \underbrace{G^c(\epsilon + \omega) 
\Delta A(\epsilon)}_{B(\omega^\prime)} 
G^{c \dag}(\epsilon - \omega') \Delta^\dag | \chi_p^- \rangle 
\end{equation}
describing
loss processes where interference between the potential from the hole
left behind and the outgoing photoelectron 
interfere\cite{chang-lang1,chang-lang2,lang2}. The `intrinsic'
losses, finally, are describe by the satellite structure of the hole
spectral function A. The diagrams f-h plus the first-order
satellite in $A$ ($c$)
give the first-order contribution to the first plasmon satellite as
obtained by Inglesfield\cite{ingles83} in the case of photoemission from core
levels.

We now turn to the first-order diagrams (d) and (e) with no interaction
lines joining the (+) and (-) parts.
In this case the interaction lines represent time-ordered correlations,
\begin{equation}
\label{eq:noloss_dh}
J_p^{(d)}  = 2 \pi \int_{-\infty}^0 d\epsilon \; \int_0^\infty \; d\omega^\prime
\delta(\epsilon + \omega - \epsilon_p)
\langle \chi_p^- | \Lambda^{(1)}(\epsilon + \omega, \epsilon)
A(\epsilon) \Delta^\dag | \chi_p^- \rangle 
\end{equation}
where
\begin{equation}
\langle n | \Lambda^{(1)}(\epsilon,\epsilon') | m \rangle =
i \int \; \frac{d \omega'}{2 \pi}
\langle n | \underbrace{G^c(\epsilon - \omega') \Delta G^c(\epsilon' - \omega')}_{
W^c(\omega')} | m \rangle
\end{equation}
is a first-order correction to the usual (time-ordered) vertex
function.  The diagram (e) gives the complex conjugate $[J_p^{(d)}]^*$.
The appearance of $\delta(\epsilon + \omega - \epsilon_p)$ in Eq. (\ref{eq:noloss_dh})
shows that they contributes to the no-loss current.

The remaining two diagrams ($k$) and ($l$), finally, describe a screening
of the photon field in the solid\cite{feibelman}. The screening will
modify the electromagnetic field by local-field effects which are
lattice-periodic deep inside the solid but more complicated near the surface.
Most photoemission calculations to date do not take these effects into account.
Effects of such screening has been considered by Liebsch\cite{liebsch3}, 
Schattke\cite{schattke1, schattke2}, and by others.
In the following
discussion they will be considered included in the optical vertex $\Lambda$.

With independent-electron propagators, the photoelectron orbital is approximated
to zeroth order in the interaction and is undamped inside the solid.
If we regard the diagrams skeletons and use full Green's functions (diagram ($c$)
should then be omitted), the photoelectron orbital gets damped inside as it should.

\subsection{Treatment based on scattering theory}
Scattering theory furnishes an equivalent description of the photoemission
process and has been the basis for several interesting new 
results\cite{hedin98,hedin99}. The obvious generalisation of the
one-electron result is to replace the one-electron scattering
orbital $ | \chi^-_p \rangle $ by its many-body counterpart
$| f \rangle \equiv | \chi^-_p,s^- \rangle \equiv | N-1, s, p, - \rangle$ 
of energy $E_f = \epsilon_p + E_s(N-1)$ with an asymptotically 
free electron $p$ and a $N-1$-electron system left behind in some 
excited state $s$,
\begin{equation}
\label{eq:jp_gold}
J_p = 2 \pi \sum_s | \langle f | \delta H | N \rangle |^2 
\delta ( E_i - E_f ))
\end{equation}
The equivalence with the response formulation can be demonstrated with
a variety of methods. One way is to consider the limiting behaviour of
the amplitude
\begin{displaymath}
\langle N - 1, s | c_p\delta H \frac{1}{E_i - H + i \eta}  | N \rangle
\end{displaymath}
inherent in the response formulation and verify that it tends to
\begin{displaymath}
\frac{1}{E_i - E_f + i \eta} \langle f | \delta H | N \rangle
\end{displaymath}
when $\eta \rightarrow 0$. There are some subtleties involved because
the photoelectron is identical the electrons in the sample.
For a discussion, see \textit{e.g.} Ref. \cite{alm85}.

A drawback with the above treatment is that every possible final many-body state
has to be treated explicitly, a clearly impossible task in a `full' treatment.
In practice, however, one has so far only been able to account for losses
to low orders. In sp-bonded materials, for example, the important
excitations are boson-like and consist of plasmons and particle-hole pairs. The 
plasmons carry the major oscillator strength. Going back to Fig. (\ref{pe1}),
diagrams ($c$) and ($f-h$) all have a real boson in the final state. By summing
over possible intermediate virtual states with some subspace, the propagators
become renormalised and one obtains an approximation to the first boson loss
satellite with damping properly included. The idea to work in subspaces
is the essence of the Hedin-Bardyszewski approach\cite{hedin85}. To make
the problem more clear-cut, Hedin following an idea of McMullen 
\textit{et al.}\cite{birger}
considered the photoelectron as distinguishable from other electrons
in the sample, the `blue-electron model'.

The very existence of the Golden-Rule expression in Eq. (\ref{eq:jp_gold})
suggests that also the results from response theory can be put in a similar
form. This is indeed the case, and a technique how it may be achieved will
be discussed at the end of this paper.

\section{No-loss current}
\label{no_loss}
Photoemission spectra typically consist of a quasi particle-like
peak and various losses at lower kinetic energy. A mathematically
precise definition of a no-loss part can only be given in the
uninteresting case where the sample ends up in the $N-1$-particle
ground state. If we are willing to accept a certain fuzziness,
we can define a "no-loss" part with about the same precision
as quasi particles below the Fermi level. In practice there is
usually no difficulty to identify, say, core-electron quasi particles,
or valence-electron quasi particles well below the Fermi energy
so long as their lifetime broadening is small on the energy scale
of interest.

\begin{figure}[htbp]
\begin{center}
\includegraphics*[scale=0.7]{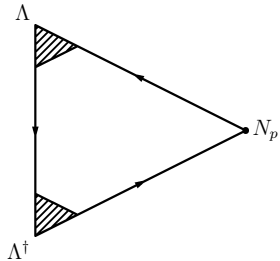}
\caption{The no-loss part of the photocurrent}
\label{noloss}
\end{center}
\end{figure}

Our foregoing analysis suggest that interaction lines which join
the positive and negative time-ordered parts of a diagram
correspond to energy losses in the final state. There must always
be at least one $G^<$ line joining the two optical vertices in
a diagram, \textit{i.e}, there is always an excitation where an
electron has been removed. In first order, the diagrams ($c$, $f-h$)
in Fig. (\ref{pe1})
contribute to additional losses involving particle-hole
pairs or plasmons because they involve $W^>$.
In an interaction-picture representation for 3-current correlation $R_p$
we have
\begin{equation}
 R_p (t,t') = \langle 0 | T^\dag \left [ S(-\infty, \infty) \delta H(t)
 c (0) \right ] T \left [ S(-\infty, \infty) c (0)\delta H(t) \right ] 
 |0\rangle .
\end{equation}
We insert a complete set of final $N-1$-electron states between the
forward- and backward time-ordered parts and expand. The $N-1$ electron
ground state corresponds to ejecting an electron from the Fermi surface
adiabatically. If the sample has well-defined quasi particles it is
natural to filter out all final states with energies a couple of lifetime
widths away for the quasi-particle energy. This corresponds essentially
to excluding interaction and electron lines which cross the two
time-ordering parts. If we sum all diagrams of this kind, we obtain
\begin{equation}
\label{eq:noloss}
J_p = 
2 \pi \; \int_{-\infty}^\mu \delta(\epsilon_p - \epsilon - \omega)
\langle \chi_p^- | \Lambda(\epsilon + \omega, \epsilon) 
A(\epsilon) \Lambda^\dag(\epsilon + \omega, \epsilon) |  \chi_p^- \rangle
\; d\epsilon 
\end{equation}
(see Fig (\ref{noloss})). Here, $\Lambda(\epsilon, \epsilon^\prime)$ is the screened,
time-ordered vector-coupling vertex.
In order to obtain only the no-loss part the integration with
respect to $\epsilon$ should be confined to the quasi particle part of $A(\epsilon)$. 
How this should be done cannot be specified exactly, as a consequence of the fact
that the sample is generally left in an unstable excited state. 
The impreciseness  is of the order of the lifetime width of the hole left behind.
Experimental angular-resolved spectra show that the quasi-particle
concept is useful also for states far below the Fermi energy and that
spectra of real systems can be considered composed of a no-loss part
that reflects the quasi-particle bandstructure, and a loss part that
reflects the dynamical response of the system.

The vertex correction is clearly related to the problem of how the optical
matrix elements should be calculated. In the dipole approximation,
the velocity ($\Delta_{kl} \sim \langle k | \vA \cdot \vp | l \rangle$), 
length ($\Delta_{kl} \sim \langle k | \vE \cdot \vrr | l \rangle$),
and acceleration ($\Delta_{kl} \sim \langle k | A \cdot \nabla w(\vrr) | l \rangle$)
formulas are equivalent. For independent particles the equivalence
readily follows by commuting the one-electron Hamiltonian with the momentum,
and $w$ is then the effective one-electron lattice potential.
When the independent-electron
propagators are replaced by their interacting counterparts, however, the
equivalence is lost. There is also a question how the effective potential
$w$ should be chosen. This problem resolved by the present author
in the mid-eighties\cite{alm86}.
The replacement of the bare
optical matrix element by the dressed vertex $\Lambda_{v/L/a}$ restores
the equivalence almost completely\cite{alm86}. The remaining discrepancy
is of order of the lifetime width of the hole left behind and is clearly
related to the fuzziness of quasi particles away from the Fermi surface.
This again shows that Eq. (\ref{eq:noloss}) furnishes a good description
in systems with well-defined quasi particles. The acceleration
matrix elements involve $\epsilon^{-1}(\omega) \vA \cdot \nabla w_{nuc}(\vrr)$
(a polarisation where are nuclei have been shifted a distance $\vA$.
The last result was first obtained by Hermeking\cite{hermeking}.

\section{Energy Losses}
\label{sec_losses}

We shall here mainly be concerned with plasmon losses in sp-bonded materials.
As is well known
plasmon losses can be produced either "intrinsically" in connection with the
primary photo-excitation, or "extrinsically" by the escaping photoelectron on 
its way out.  At lower kinetic energies the two modes
of plasmon excitation interfere, leading to a suppression of the
satellite
strength. This interference has in the past been described using
semi-classical models where the photoelectron is assumed to follow a classical
trajectory and where the photoelectron and the hole left behind are represented
by external potentials\cite{mahan73,gadzuk75,longe,tougaard}.
In the case of core-electron photoemission, 
Hedin \textit{et al.}\cite{hedin98,hedin2003} have
shown that a fully quantum-mechanical description of the leading
satellites agrees quite well with  a semi-classical description for excitation
energies a couple of plasmon energies above threshold. It also shows that
interference between intrinsic and extrinsic plasmon production persists
up to the keV range, thereby confirming the earlier works by
Chang and Langreth\cite{chang-lang1, chang-lang2}. However, there are
to our knowledge no similar studies for photoemission from valence states.

As will be discussed below, a straight-forward expansion in orders of the screened
interaction $W$ will not in general guarantee positive spectral densities.
I will here outline a different summation order in  which the partial sums
have a golden-rule-like form involving transition amplitudes $\tau_{fi}$. 
Thus, each partial sum has the
generic form
\begin{equation}
I(\omega) = \sum_f |\tau_{fi}|^2 \delta(E_i + \omega - I_f),
\label{eq:GR}
\end{equation}
where $i, f$ label real excitations in the initial and final states, 
respectively.  The transition amplitudes can be related to cut Keldysh diagram.
By performing an expansion of the amplitudes to a given order, each partial sum
gives a positive result.
The procedure amounts to splitting complete Keldysh
diagrams in pieces and to sum them a particular order such that every
partial sum is positive. If carried out of infinite order, `all' complete
diagrams are summed. I will outline the procedure with the low-order plasmon 
losses in photoemission as an example, and give some new model 
results for valence photoemission from a simple-metal surface.

\subsection{Lowest-order contributions}
\label{sec:loss1}

Let us begin with the lowest-order contributions where no real
losses have occurred. Up to first order in $W$,
the zeroth-order diagram in Fig. \ref{pe0}
and diagrams (d) and (e) in Fig. \ref{pe1} contribute
and will involve the combination
\begin{displaymath}
\langle \chi_p^- | \Delta A(\epsilon) \Delta^\dag | \chi_p^- \rangle +
2 \Re \langle \chi_p^- | \Lambda^{(1)}(\epsilon + \omega, \epsilon)
\Delta A(\epsilon) \Delta^\dag | \chi_p^- \rangle .
\end{displaymath}
The obvious missing piece in the above approximation is the term
$\langle \chi_p^- | \Lambda^{(1)} A(\epsilon) \Lambda^{(1)\dag} | \chi_p^- \rangle$.
If this term is added, a manifestly positive result is obtained,
\begin{displaymath}
\langle \chi_p^- | [ \Delta + \Lambda^{(1)}] 
A(\epsilon) [\Delta + \Lambda^{(1)}]^\dag | \chi_p^- \rangle 
\end{displaymath}
in which the transition amplitudes rather than the spectral intensity
has been truncated to a given order.

Contributions with no real excitations in the final state are somewhat
artificial and will give no contribution when particle-hole
excitations are included in the screened interaction. The mechanism 
is essentially the same as in the Langreth theorem\cite{lang3}. 
In the case of the simpler $G^<$ function we have
\begin{equation}
G^<(\epsilon) = [1 + G^r (\epsilon) \Sigma^r(\epsilon)] G_0^< (\epsilon)
[1 + \Sigma^a(\epsilon) G^a(\epsilon)] + G^r (\epsilon)
\Sigma^<(\epsilon) G^a(\epsilon) .
\label{eq:langtheo}
\end{equation}
If there are no real excitations in a particular spectral region,
$\Sigma^<$ vanishes, and the first term in Eq. (\ref{eq:langtheo})
gives the entire contribution. As soon as real excitations occur,
the first term tends to zero in the infinite-volume limit and the
entire contribution comes from the second term. What actually occurs
is that previously sharp single-particle-like excitations become
broadened. By the same mechanism, contributions with no real
excitations in the final state except a perfectly sharp quasi particle
will not contribute to the photoemission current when broadening by
particle-hole pairs is included.

We next proceed to the first-order
diagrams $(c)$ and $(f-h)$ in Fig (\ref{pe1}) which contribute to
a one-boson loss.
These four diagrams nicely combine into a complete
square of intrinsic and extrinsic satellite amplitudes.
We make use of the fact that the spectral functions $B(\omega)$
and $A(\epsilon)$ are Hermitian and positive-definite.
They can thus be diagonalised,
\begin{eqnarray}
\label{eq:diagB}
\langle \vrr | B(\omega) | \vrr' \rangle &=& \sum_\mu b_\mu^2(\omega)
v_\mu(\omega, \vrr) v_\mu^*(\omega, \vrr') , \\
\label{eq:diagA}
\langle \vrr | A(\epsilon) | \vrr' \rangle &=& \sum_n a_n^2(\epsilon)
u_n(\epsilon, \vrr) u_n^*(\epsilon, \vrr') .
\end{eqnarray}

The ``intrinsic' part $(c)$ can be written
\begin{equation}
\label{eq:loss_c}
J_p^{(c)} = 2 \pi 
\langle \chi_p^- | 
 \Delta A^<(\epsilon_p - \omega) \Delta^\dag 
| \chi_p^- \rangle ,
\end{equation}
where $A^< = G^c \Gamma^< G^{c \dag}$ and $\Gamma^< = - i \Sigma^< / 2 \pi$.
In the $GW$ approximation\cite{hedin65}, we have
\begin{eqnarray}
\lefteqn{\langle \vrr | \Gamma^<(\epsilon_p - \omega) | \vrr' 
\rangle =}\nonumber \\
& & \sum_{n\mu} \int d \omega ' a^2_n (\epsilon_p + \omega ' - \omega) b^2_\mu
( \omega')
u_n(\epsilon_p + \omega ' - \omega, \vrr) u_n^*(\epsilon_p + \omega ' - \omega, \vrr')
 v_\mu(\omega', \vrr ) v_\mu^*(\omega', \vrr )\nonumber \; .
\end{eqnarray}
We insert this in Eq. (\ref{eq:loss_f}) to obtain
\begin{equation}
J_p^{(c)} = 2 \pi \sum_{n\mu} \int_0^\infty d \omega'  | 
M_{1,1,a}^{n\mu}(\omega') |^2
a_n^2(\epsilon_p + \omega' - \omega) b_\mu^2(\omega') \; ,
\end{equation}
where 
\begin{equation}
\label{eq:M1}
M_{1,1,a}^{n\mu} = \langle \chi_p^- | \Delta G^c(\epsilon_p - \omega)
| v_\mu^*(\omega') u_n(\epsilon_p + \omega' - \omega) \rangle \; .
\end{equation}
We next turn to the purely extrinsic part ($f$). Again we rewrite
$A^<$ and $B(\omega)$ in a diagonalising basis to obtain
\begin{equation}
J_p^{(f)} = 2 \pi \sum_{n\mu} \int_0^\infty d \omega'  | M_{1,1,b}^{n\mu}
(\omega') |^2
a_n^2(\epsilon_p + \omega' - \omega) b_\mu^2(\omega') ,
\end{equation}
where in this case
\begin{equation}
\label{eq:M2}
M_{1,1,b}^{n\mu} =
\langle \chi_p^- v_\mu(\omega') | G^c(\epsilon_p + \omega')
\Delta | u_n(\epsilon_p + \omega' - \omega) \rangle .
\end{equation}
The two interference terms $(g-h)$, finally, can be expressed
in a similar way but with matrix elements $M_{1,1,a}^* M_{1,1,b}$.
The total contribution can then be written
\begin{equation}
J_p^{(1,1)} = 2 \pi \sum_{n\mu} \int_0^\infty d \omega'
\left | M_{1,1}^{n\mu}(\omega') \right |^2
a_n^2(\epsilon_p + \omega' - \omega) b_m^2(\omega') \; ,
\end{equation}
where
\begin{equation}
M_{1,1} = M_{1,1,a} + M_{1,1,b} \;.
\end{equation}
The approximation of keeping only the renormalised diagrams
($c, f-h$) has a close correspondence to the $GW$ approximation
and may be termed a `$GW$ approximation for photoemission'.

In the simplest non-selfconsistent $G_0W$ 
approximation, the spectral function in $G^<$ should be approximated by its
independent-particle equivalent in $\Sigma^\lessgtr$. (For different levels
of selfconsistency, see Ref. \cite{holm_barth98}.) In this
approximation,
$a_m^2(\epsilon)$ is replaced by $\delta(\epsilon - \epsilon_m)$ in the
above expressions, and remaining propagators by their $G_0W$ equivalents.
One can also in principle
do partial selfconsistency with respect to $G$, which in this
context amounts to calculate $A(\epsilon)$ selfconsistently
from Eq. (\ref{eq:diagA}), the resulting $\Sigma^\lessgtr$, and the
Langreth theorem, Eq. (\ref{eq:langtheo}).

\begin{figure}[htbp]
\begin{center}
\includegraphics*[scale=0.65]{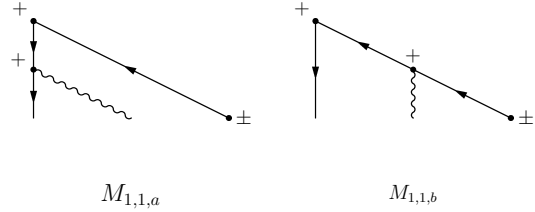}
\caption{Transition amplitudes to first order as cut Keldysh diagrams }
\label{fig_m1}
\end{center}
\end{figure}

The matrix elements $M_{1,1,a}$ and $M_{1,1,b}$ can be represented graphically
at cut diagrams with only + vertices and one dangling interaction line
$W^>$ as in Fig. \ref{fig_m1}.
When joined with the corresponding lower
half, the four diagrams (c), (f-h) will emerge. 

\subsection{Structure of higher-order contributions}
\label{sec_loss2}

Already the above $GW$-like approximation gives at least a qualitative
description of how the leading plasmon satellite is modified by interference.
However, although $GW$ theory is usually quite accurate for quasi particle
positions it gives a somewhat poor representation of the satellite shape.
In the spectral function $A(\epsilon)$ there is only one satellite
which is too broad and with a mean energy too far away from the quasi particle.
By adding the first vertex diagram and self-consistency diagrams one improves
the satellite in the case of core electrons, and model studies indicate
that also the valence electron spectrum is improved. A corresponding
approximation for photoemission will involve at least two screened interactions.

\begin{figure}[htbp]
\begin{center}
\includegraphics*[scale=0.65]{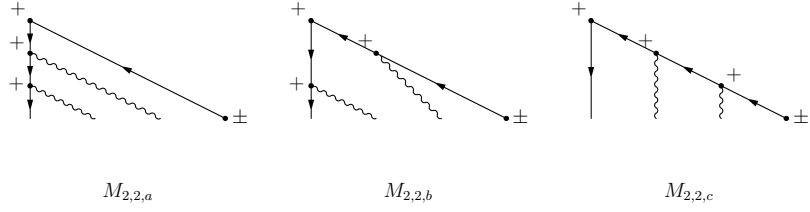}
\caption{Transition amplitudes contributing to the second plasmon satellite }
\label{fig_m2}
\end{center}
\end{figure} 

\begin{figure}[htbp]
\begin{center}
\includegraphics*[scale=0.65]{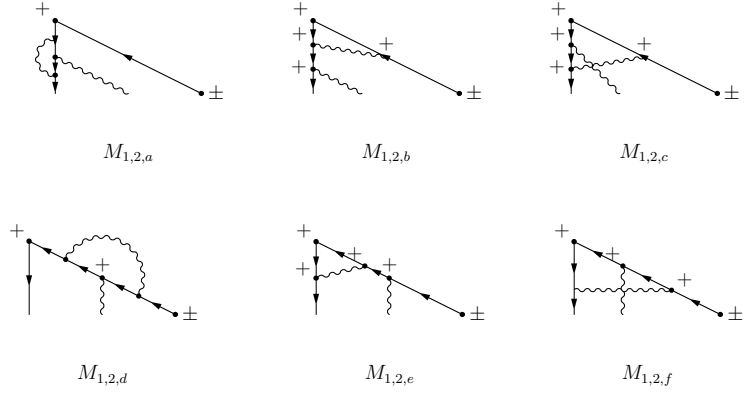}
\caption{Transition amplitudes modifying the first plasmon satellite }
\label{fig_m1b}
\end{center}
\end{figure} 

Starting with second order theory, a large number of diagrams occurs
which most probably gives very small contributions. The screening
of the photons can now in principle interfere with the excitations
shaken up by the photoelectron. If such contributions are neglected,
the coupling to the photon field enters solely in form of a time-ordered
screened optical vertex $\Lambda$ as in Eq. (\ref{eq:noloss}). 
Except possibly very close to threshold, the important processes are those
where photoelectron, once created, remains until it is detected. In terms
of diagrams this corresponds to adding screened interaction lines to the
basic independent-electron triangle in Fig. \ref{pe0}. The leading-order 
amplitudes which contribute to the second plasmon satellite are illustrated 
in Fig. \ref{fig_m2}. The dangling interaction lines can be combined in two
different ways when complete photoemission diagrams are formed. As a
result, the amplitudes in Fig. \ref{fig_m2} produce 18 complete diagrams.
The contribution from these 18 diagrams can be written
\begin{eqnarray}
J_p^{(2,2)} &=& 2 \pi \sum_{n\mu\nu} \int_0^\infty d \omega_1 d \omega_2
\left | \sum_{l =  a, b, c} M_{2,2,l}^{n\mu \nu} + 
M_{2,2,l}^{n\nu \mu} \right |^2
\nonumber \\
&\times&
a_n^2(\epsilon_p + \omega_1 + \omega_2 - \omega)
b_\mu^2(\omega_1)b_\nu^2(\omega_2) ,
\end{eqnarray}
where
\begin{eqnarray}
M_{2,2,a}^{n\mu \nu} &=& \langle  \chi_p^- | \Delta G^c(\epsilon_p - \omega)
v_\mu(\omega_1) G^c(\epsilon_p + \omega_1 - \omega) |
v_\nu(\omega_2) u_n(\epsilon_p + \omega_1 + \omega_2 - \omega) \rangle
\nonumber \\
M_{2,2,b}^{n\mu \nu} &=& \langle  \chi_p^- v_\mu(\omega_1) | G^c(\epsilon_p + \omega_1) \Delta 
G^c(\epsilon_p + \omega_1 - \omega) |
v_\nu(\omega_2) u_n(\epsilon_p + \omega_1 + \omega_2 - \omega) \rangle
\\
M_{2,2,c}^{n\mu \nu} &=& \langle  \chi_p^- v_\mu(\omega_1) | G^c(\epsilon_p + \omega_1) 
v_\nu(\omega_2) G^c(\epsilon_p + \omega_1 + \omega_2) \Delta |
v_\nu(\omega_2) u_n(\epsilon_p + \omega_1 + \omega_2 - \omega) \rangle
\nonumber
\end{eqnarray}

In addition, there are amplitudes which modify the
shape of the first plasmon satellite containing one dangling interaction
line (Fig. \ref{fig_m1b}).
The latter amplitudes are to be added to
the first-order amplitudes which are the squared and summed. The diagrams
which result from pairing off with the amplitudes $M_{1,1}$ in Fig. \ref{fig_m1}
are of second order in $W$. In this way 24 complete Keldysh diagrams are 
obtained, and they give a contribution of the form
\begin{eqnarray}
J_p^{(1,2a)} &=& 4 \pi \Re \sum_{n\mu} \int_0^\infty d \omega_1 
 \left [ M_{1,1}^{n\mu} \right ]^*
M_{1,2}^{n\mu} \;
a_n^2(\epsilon_p + \omega_1 - \omega)
b_\mu^2(\omega_1) \; ,
\label{eq:j12a}
\end{eqnarray}
where $M_{1,2}$ is the sum of the six amplitudes $M_{1,2,l}$ in Fig.
\ref{fig_m1b}.
The sum
$J_p^{(1,1)} + J_p^{(1,2a)}$
and thereby the complete second-order current 
$J_p^{(1,1)} + J_p^{(1,2a)} + J_p^{(2,2)}$ 
is not positive-definite. If we add the diagrams where the second-order 
amplitudes to the first boson satellite are paired against 
themselves a manifestly positive 
contribution to the first boson satellite is obtained,
\begin{eqnarray}
J_p^{(1,2)} &=& 2 \pi \sum_{n\mu} \int_0^\infty d \omega_1 
\left | M_{1,1}^{n\mu} + M_{1,2}^{n\mu} \right |^2
a_n^2(\epsilon_p + \omega_1 - \omega)
b_\mu^2(\omega_1) .
\label{eq:j12}
\end{eqnarray}
The sum $J_p^{(1,2)} + J_p^{(2,2)}$ includes all second-order and part of
the third-order contributions to the photoelectron current.

The above procedure can evidently be extended to any order. The basic idea
is to sum transition amplitudes to a given order, and then form all diagrams
containing these amplitudes by pairing as above. For a given order $n$
the current is the sum $J^{(1,n)}_p + J^{(2,n)}_p + \cdots + J^{(n,n)}_p$ 
corresponding to $1, 2, \ldots, n$ boson losses. Each partial current
$J^{(r,n)}_p$ will involve amplitudes with $r$ dangling and
up to $n-r$ time-ordered interaction lines. The partial current
$J^{(r,n)}_p$ is obtained by pairing these amplitudes in all
possible ways. The partial current will include all diagrams
up to order $n$, and part of the contribution of order between
$n$ and $2n-r$ and it will be manifestly positive.
If one could continue indefinitely, eventually all diagrams
will be included. The method of expanding transition amplitudes rather than spectral
densities is also applicable for obtaining approximations
to the one-electron spectral function and to the dynamical structure
factor.

\subsection{Model calculations of plasmon and electron-hole losses}

I here describe model calculations in the simplest non-trivial
approximation in which we keep the four
diagrams $(c)$ and $(f-h)$ in Fig (\ref{pe1}).

In the expressions for matrix elements, $G^c$ one can usually
approximate it by $G^r$ in Eq. (\ref{eq:M2}) and by a $G^a$
in Eq. (\ref{eq:M1}).
In our model calculations we have further approximated $G^{r/a}$ by 
$1/(h_0 \pm i \Gamma)$
with a damping from the on-shell value of $\Im \Sigma$. The matrix
elements can then be evaluated by propagating the orbitals in space
under influence of a non-Hermitian Hamiltonian which greatly
simplifies the calculations. The system was approximated by a
a semi-infinite jellium perturbed by a weak lattice
pseudo potential. (Without a lattice potential only surface
photoemission can occur because of momentum conservation.)
The dielectric function and thus $B(\omega)$ was approximated
by a semi-analytical result by Bechstedt 
\textit{et al.}\cite{bechstedt}.

\begin{figure}[htbp]
\begin{center}
\includegraphics*[angle=270,scale=0.35]{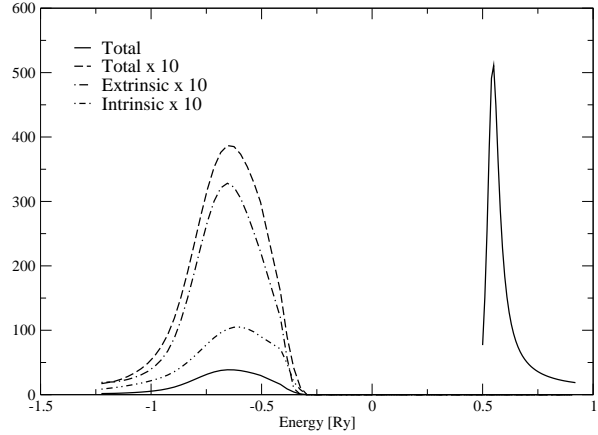}
\caption{Photocurrent as function of photoelectron energy in Ry. 
The zero level corresponds to the bottom of the valence band. 
The spectrum is excited 7.5 Ry above threshold, and the photoelectron 
mean free path is approximated by a constant corresponding to 
$\Gamma_p = 0.3$ Ry.}
\label{pl7.5}
\end{center}
\end{figure}

\begin{figure}[htbp]
\begin{center}
\includegraphics*[angle=270,scale=0.35]{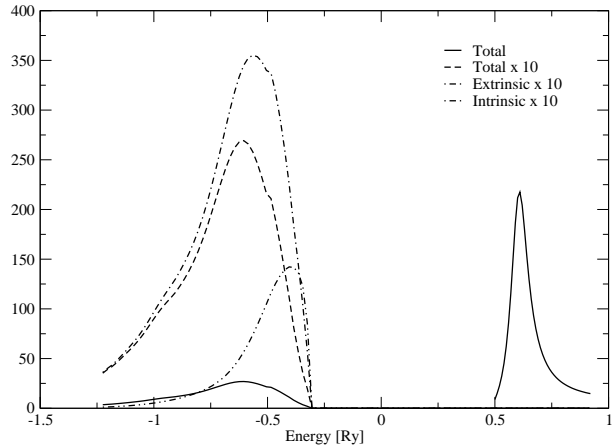}
\caption{Photocurrent excited 4.5 Ry above threshold. $\Gamma_p = 0.3$ Ry. 
Notations as in Fig. \ref{pl7.5}}
\label{pla4.5}
\end{center}
\end{figure}

In Fig. (\ref{pl7.5}) we show results for plasmon losses
with an excitation energy 7.5 Ry above threshold. These
and the following results correspond to aluminium, and 
the lattice $\vG$ vector, which acts as momentum source, has a parallel
component of 0.7 a.u. and a component of 2.6 a.u. perpendicular 
to the surface.
The extrinsic losses dominate, and the total loss is
somewhat smaller than the sum of the extrinsic and intrinsic
ones. However, we have observed that interference
depend on the direction of the photocurrent even at 7.5 Ry. The two
loss mechanisms are therefore not completely decoupled
even at \~ 100 eV. In Fig. (\ref{pla4.5}) we have lowered
the excitation energy to 4.5 Ry. In this case the interference
is much stronger, and the extrinsic and intrinsic parts
interfere destructively.


\begin{figure}[htbp]
\begin{center}
\includegraphics*[angle=270,scale=0.35]{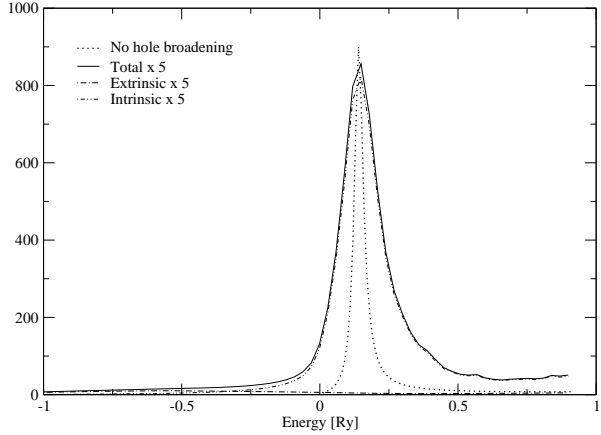}
\caption{Broadening of the quasi particle peak at normal emission. 
$\omega_i = 2.5$ Ry.
For notations see Fig. \ref{pl7.5}}
\label{peh2.5}
\end{center}
\end{figure}

Finally one may ask how the quasi particle line shape is modified
by the transport and interference effects. In our calculations
we have obtained almost negligible effects.
As an example, I show 
the quasi particle region at an excitation energy of 2.5 Ry
(Fig. (\ref{peh2.5})).
The broadening that is seen without
any particle-hole effects comes from the
finite mean free path which smears the momentum selection
rules normal to the surface.

As remarked above, the first-order $GW$ theory gives a rather crude
representation of the shape of the plasmon satellites. 
The description will be improved by going to second order in transition
amplitudes as described in Sec. \ref{sec_loss2}.
This higher-order approximation is still simple enough to
allow for numerical evaluations.
Within the approximations used above for the first-order results
the second-order amplitudes can again be obtained by 
propagating one-electron orbitals which makes the calculation manageable. 
As the excitation energy is further increased, more and more plasmon satellites
will result in experimental spectra. At sufficiently high energies,
the energy losses can be described by semi-classical approaches in which
the photoelectron is considered as a classical particle which moves
along a certain trajectory\cite{ah83,hedin2003} or by an integral equation
which approximates the losses at high energies\cite{chang-lang2}. We expect
that explicit quantum-mechanical calculations up to second order in $W$ will
cover the energy window up to the limit where these simpler methods are applicable.
Extensions to second order along the lines given here are on the way
and will be presented elsewhere.

\section{Concluding remarks}

In this paper I have discussed how
techniques from non-equilibrium Green's function theory may be used
to develop useful approximations as well as for establishing
general results such as the structure of the no-loss quasi particle
part of the spectrum. Although I have discussed energy losses in
some detail in the second part, I would like to emphasis that it
is the no-loss quasi particle part that is usually of primary
interest in experimental studies. The current computational schemes
in which the photoelectron moves in a non-Hermitian optical potential
is essentially correct. The necessary ingredients are here sufficiently
accurate approximations for the self-energy including its imaginary
part. It may be necessary to account for its non-locality at least
in some approximate way. The proper evaluation of matrix elements
the screened optical field is another area where improvements
may be necessary. Especially in spectral regions where the photoelectron
self-energy is rapidly varying it may also be necessary to account
for the vertex corrections discussed in Sec. \ref{no_loss}.

In order to treat energy losses below the main quasi-particle peak,
a regrouping of diagrams in many-body perturbation theory has been proposed. 
In each order the partial sums have a Golden-Rule-like form which guarantees positive
spectral intensities.  This way of regrouping may be useful
also in other problems involving spectral intensities. One such
problem is how to improve the one-electron spectral function
beyond $GW$ theory.

Some new model results for valence-electron
photoemission have been presented, and more realistic results are on the way. 
The model results confirm the
expectations that interference effects modify the shape and
strength of plasmon satellites also quite far away from the
excitation threshold, whereas the quasi-particle part is little modified.
Based on our ongoing work we intend to compare fully quantum-mechanical
results to semi-classical descriptions
of the photoelectron transport away from threshold.

\section*{Acknowledgements}
It is a pleasure to thank the organisers and especially Michael Bonitz
for an excellent workshop.
The present work was supported by the EU's 6th Framework Programme through
the NANOQUANTA Network of Excellence (NMP4-CT-2004-500198).
\section*{References}


\bibliographystyle{unsrt}
\bibliography{kiel}

\end{document}